\documentclass{emulateapj}

\shorttitle{Black Hole in BW Cir}
\shortauthors{J. Casares et al.}

\begin{document}

\title{Evidence for a Black Hole in the X-ray transient GS 1354-64 (=BW Cir)}

\author{J. Casares\altaffilmark{1}, C. Zurita\altaffilmark{2}, 
T. Shahbaz\altaffilmark{3}, P.A. Charles\altaffilmark{4} and 
R.P. Fender\altaffilmark{5}}






\altaffiltext{1}{Instituto de Astrof\'\i{}sica de Canarias, 38200 La Laguna, 
Tenerife, Spain; jcv@ll.iac.es} 
\altaffiltext{2}{Observat\'orio Astron\'omico de Lisboa,  
Tapada da Ajuda, 1349-018 Lisboa, Portugal; czurita@oal.ul.pt} 
\altaffiltext{3}{Instituto de Astrof\'\i{}sica de Canarias, 38200 La Laguna,
Tenerife, Spain; tsh@ll.iac.es}
\altaffiltext{4}{Dept of Physics \& Astronomy, University of Southampton, 
Southampton, SO17 1BJ, UK; pac@astro.soton.ac.uk}
\altaffiltext{5}{Astronomical Institute 'Anton Pannekoek', University of Amsterdam, 
Kruislaan 403, 1098 SJ Amsterdam, the Nederlands; rpf@science.uva.nl}

\begin{abstract}

We present the first radial velocity curve of the companion star to BW
Cir which demonstrates the presence of a black hole in this X-ray 
transient which recorded outbursts in 1987 and 1997 (and possibly 1971-2). 
We identify weak absorption features corresponding to
a G0-5III donor star, strongly veiled by a residual accretion disc which
contributes 61-65 \% of the total light at $\lambda$6300. The
Doppler motions of these features trace an orbit of $P=2.54448$ days
(or its 1-yr alias of $P=2.56358$ days) and velocity semi-amplitude
$K_{2}=279 \pm 5$ km s$^{-1}$ (or $K_{2}=292 \pm 5$ km s$^{-1}$). Both
solutions are equally possible.  The mass function implied by the
shorter period solution is $f(M)= 5.75 \pm 0.30 M_{\odot}$ which,
combined with the rotational broadening of the tidally locked
companion ($V \sin i =71 \pm 4$ km s$^{-1}$), yields a compact object
mass of $M_{1} \sin ^3 i = 7.34 \pm 0.46 M_{\odot}$. This is 
substantially above the mass of a neutron star under any standard equation 
of state of nuclear matter. The companion star is probably a G subgiant 
which has evolved off the main sequence in order to fill its Roche lobe.  
Remarkably, a distance of $\ge$27 kpc is inferred by the companion's 
luminosity and this is supported by the large observed systemic velocity 
($\gamma = 103 \pm 4$ km s$^{-1}$) which requires such a distance in order 
to be consistent with the Galactic rotation curve.  

\end{abstract}


\section{Introduction}
Low mass X-ray binaries (LMXBs) are interacting binaries where a low
mass star transfers matter onto a neutron star or black hole (e.g. Charles 
\& Coe 2004). Mass transfer takes place through an
accretion disc where angular momentum is removed and gravitational
potential energy converted into (high-energy) radiation. Accretion
discs evolve toward stationary states where the mass transfer rate
through the disc, $\dot{M}$, adjusts towards the value of $\dot{M_2}$,
the mass transfer rate supplied by the donor star. $\dot{M_2}$ is
driven by the binary/donor evolution and if $\dot{M_2} <
\dot{M}_{crit} \sim 10^{-9}$ M$_{\odot}$ yr$^{-1}$ then mass transfer
instability cycles (outbursts) are triggered (van Paradijs 1996). There 
are $\sim$200 bright (L$_{X} \simeq 10^{36}-10^{38}$ erg
s$^{-1}$) LMXBs in the Galaxy and most of them harbour neutron stars as
implied by the detection of X-ray bursts/pulsations. On the other hand
$>70$ \% of transient LMXBs are known to harbour accreting black
holes, as demonstrated by dynamical studies of the faint companion
stars, which is mainly possible when X-rays switch off (but see Hynes 
et al. 2003). We currently have dynamical evidence for 16 black hole LMXBs, 
with mass functions ranging from $0.22\pm0.02$ M$_{\odot}$ up to $9.5\pm 
3.0$ M$_{\odot}$ (see Charles \& Coe 2004, McClintock \& Remillard 2004).  
The mass spectrum of black holes is of crucial astrophysical significance 
in constraining supernovae models and the equation of state of nuclear
matter. Clearly both more, and more accurate, black hole mass
determinations are required before these issues can be fully
addressed. 

BW Cir is the optical counterpart of the X-ray transient GS 1354-64,
discovered in 1987 by the Ginga satellite (Makino et al. 1987). 
It displayed X-ray properties reminiscent of BH transients
i.e. a combination of a soft multi-BB component, with inner disc
temperature of $\sim$0.7 keV, plus a hard power-law tail with photon
index 2.1 (Kitamoto et al. 1990). BW Cir went into outburst again in
1997 but remained in a low/hard state throughout (Revnivtsev et al. 2000, 
Brocksopp et al. 2001). Interestingly, its sky position coincides with 
two older recorded X-ray
transients, Cen X-2 and MX 1353-64, discovered in 1966 and
1971-2 respectively (see Kitamoto et al.  1990). The former was the
first X-ray transient ever discovered, with a peak soft X-ray
luminosity of $\sim$ 8 Crab. If this activity were attributed to the
same source, then it would make BW Cir a BH transient with one of the 
shortest recurrence time ($\simeq$ 8-10 years). However, the X-ray 
properties of these events were markedly different and hence, if 
caused by the same source it would indicate that it has displayed at 
least four distinct X-ray states. 

BW Cir settles down in quiescence at R=20.5 (Martin 1995) and as yet no 
indication of the orbital period exists. A possible period of $\sim$ 46 
hr and 0.3-0.4 V-mag amplitude was reported during the 1987 outburst 
(Ilovaisky et al. 1987) but it has not been confirmed. On the other hand, 
Martin (1995) suggests a very tentative 15.6 hr periodicity with $\sim$0.1 
R-mag amplitude from quiescent data although the folded light 
curve does not mimic the classic ellipsoidal modulation. 
Here we present the first spectroscopic detection of the companion star in 
BW Cir and the analysis of its radial velocity curve. This provides the first
dynamical probe for the presence of a black hole in this historical transient 
X-ray binary. 

\section{Observations and Data Reduction} 

We have observed BW Cir using the FORS2 Spectrograph attached to the
8.2m Yepun Telescope (UT4) at Observatorio Monte Paranal (ESO) on the
nights of 22-23 June 2003, 14-15 and 25-27 May 2004.  A total of 55
spectra were collected with integration times varying between 1800 and
2000s (depending on atmospheric conditions).  The R1200R holographic
grating was employed which, combined with a 0.7 arcsec slit, produced
a wavelength coverage of $\lambda\lambda$5745-7230 at 70 km s$^{-1}$
(FWHM) resolution, as measured from gaussian fits to the arc lines.  A
He+Ne+Hg+Cd comparison lamp image was obtained with the telescope in
park position to provide the wavelength calibration scale. This was
derived by a 4th-order polynomial fit to 37 lines, resulting in a
dispersion of 0.74 \AA~pix$^{-1}$ and an rms scatter of 0.04
\AA. Instrumental flexure in our target spectra was monitored through 
cross-correlation
between sky spectra and it was always less than 33 km s$^{-1}$.  These
velocity drifts were removed from each individual spectrum, and the
zero point of the final wavelength scale was fixed to the strong OI
sky line at $\lambda$6300.304.  Two additional 1800s spectra were
obtained on the nights of 23-24 Aug 2000.  Here we used the R600R
grism in combination with a 0.7 arcsec slit which resulted in a resolution 
a factor of two lower.

We also observed several radial velocity and spectral type standards with 
exactly the same instrumental configuration on the nights of 17 May 2003 and 
27-28 May 2004. These cover spectral types F5III-K7III. 
All the images were de-biased and flat-fielded, and one-dimensional 
spectra extracted using conventional optimal extraction techniques in 
order to optimize the signal-to-noise ratio of the output (Horne 1986). 

\begin{figure}
\includegraphics[angle=-90, scale=0.33]{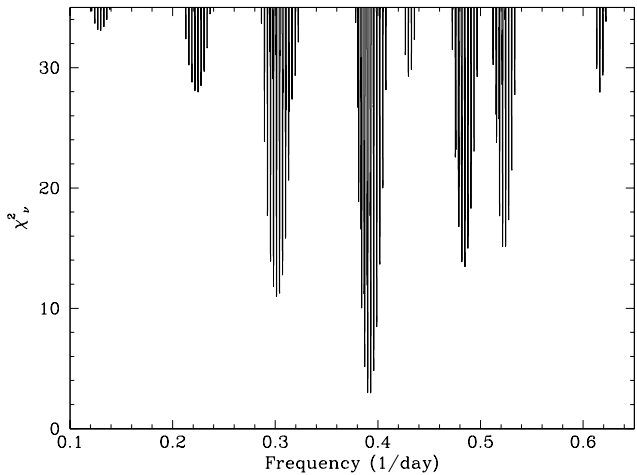}
\caption{$\chi^2$ periodogram of  the radial velocities obtained by
cross-correlation with the G5III template HD 62351. The best periods are 
at 0.393 day$^{-1}$ (2.5445 d) and its 1-year alias at 0.390 day$^{-1}$ 
(2.5641 d).  
\label{fig1}}
\end{figure}

\section{The Radial Velocity Curve and System Parameters}

The 57 individual spectra were prepared for cross-correlation analysis
by subtracting a low order spline fit to the continuum and rebinning
the wavelength scale into a constant step of 34 km s$^{-1}$ per
pixel.  Radial velocity points were extracted from our individual
target spectra by cross-correlation with the radial velocity templates
in the spectral range $\lambda\lambda$5950-7200, after masking out all the
emission, telluric and IS absorption lines. The radial velocity points
show clear night-to-night velocity changes which strongly points
towards a $\sim$ 2d orbit. This was best seen on the nights of 14-15
May 2004, when two continuous 9hr runs could be obtained. In order to
search for the orbital period we have performed a power spectrum
analysis on the radial velocities obtained with the G5III template and
the results are displayed in Figure 1.  Frequencies longer than 0.5
cycles d$^{-1}$ (periods $<$ 2 days) can be ruled out since they
provide $\chi^{2}_\nu > 15$. The minimum $\chi^{2}_{\nu}$ (=2.98)
is found at $0.3930$ d$^{-1}$, corresponding to $P=2.54446(6)$ days.
However, we cannot rule out the 1-year alias of $P=2.56358(7)$ days
with $\chi^{2}_{\nu}=3.01$. All other aliases have $\chi^{2}_{\nu}>5$ 
and can be rejected since they are not significant at the 99.99 \%.

We have performed least-squared sine-wave fits to the radial velocity curves 
obtained with all templates and find that spectral types in the range 
F5-G8III give the best fits with $\chi^{2}_{\nu}$ in the range 2-3. We 
decided to adopt the fitting parameters of the G5III template 
since this provides the closest representation to the observed spectrum 
(see below) i.e.  
\smallskip

$ \gamma= 103.1 \pm 4.2~ {\rm km s}^{-1}; P=2.54448 \pm 0.00015~ {\rm d}; 
T_{0} = 2453140.985 \pm 0.013; K_{2} = 279.3 \pm 4.9~ {\rm km s}^{-1} 
~{\rm or}$

$ \gamma= 94.8 \pm 4.2~ {\rm km s}^{-1}; P=2.56358 \pm 0.00015~ {\rm d}; 
T_{0} = 2453140.939 \pm 0.013; K_{2} = 291.7 \pm 5.3~ {\rm km s}^{-1}$

\smallskip
\noindent 
where $T_{0}$ corresponds to standard phase 0 i.e. the inferior
conjunction of the optical star. All quoted errors are 68 \%
confidence and we have rescaled the errors so that the minimum 
reduced $\chi^{2}$ is 1.0. The
$\gamma$-velocity has been corrected from the template's radial
velocity and we note that it is unusually large for black-hole
binaries, and only comparable to the long period microquasars GRO
J1655-40 and V4641 Sgr.  Figure 2 presents
the radial velocity points, folded on the 2.54 day period, with the
best sine fit superimposed.  The mass function of BW Cir, for the case
of the short period solution, is $f(M)= M_{1} \sin^{3} i / (1+q)^{2}
= P K_{2}^{3}/2 \pi G = 5.75 \pm 0.30 M_{\odot}$ , which defines a
lower limit to the compact object's mass. This is substantially above
any neutron star mass defined by all standard equations of state of
condensed matter and hence we conclude that BW Cir contains a black hole 
(Rhodes \& Ruffini 1974).

\begin{figure}
\includegraphics[angle=-90, scale=0.33]{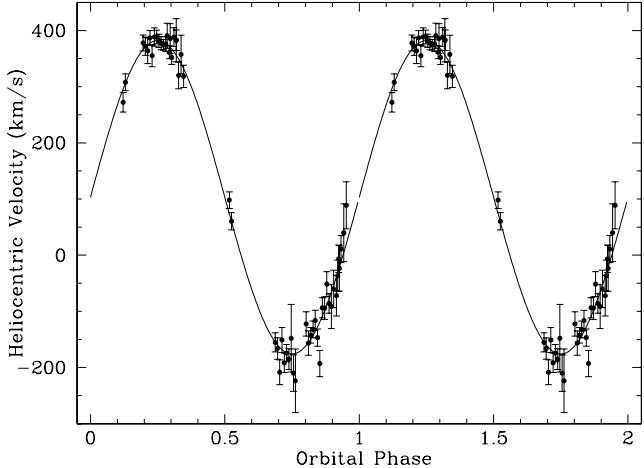}
\caption{Radial velocity points of BW Cir folded on $P=2.54448$ days and 
$T_0 = 2453140.985$ with the best fitting sinusoid.  
\label{fig2}}
\end{figure}

Furthermore, our high instrumental resolution (70 km s$^{-1}$) enables
us to measure the rotational broadening of the companion star and we can use 
this information to set a more
stringent limit to the black hole mass. We have broadened our G5III
template from 50 to 100 km s$^{-1}$ in steps of 5 km
s$^{-1}$, using a Gray profile (Gray 1992) with a linear
limb-darkening law with coefficient $\epsilon$=0.62, interpolated for
$\lambda$=6300 \AA\ and $T_{\rm eff} \simeq$ 5000 K (see Al-Naimy 1978). The broadened versions of the
template star were multiplied by fractions $f<1$, to account for the
fractional contribution to the total light, and subsequently
subtracted from the grand sum spectrum of BW Cir. The latter was
produced after averaging the 55 high resolution spectra of BW Cir in
the rest frame of the companion using the orbital solution above. A
$\chi^{2}$ test on the residuals yields $V \sin i = 71 \pm 4$ km
s$^{-1}$ (we have repeated the same experiment with all the templates
and $V \sin i$ always ranges between 59-71 km s$^{-1}$, see table
1). In the case of a tidally locked Roche-lobe filling star, $V \sin i$
relates to the $K_{2}$-velocity and the mass ratio $q=M_2/M_1$ through
the expression $V \sin i \simeq 0.462 K_2 q^{1/3} (1+q)^{2/3}$ (Wade
\& Horne 1988), from which we obtain $q=0.13 \pm 0.02$ and, therefore,
$M_1 \sin^{3} i = 7.34 \pm 0.46 M_{\odot}$. Since BW Cir does not
exhibit X-ray eclipses, the inclination angle must be $i\leq 77
^{\circ}$ (for $q=0.13$) and, consequently, the masses of the black
hole and the companion star are $M_1 \ge 7.83 \pm 0.50 M_{\odot}$
and $M_2 \ge 1.02 \pm 0.17 M_{\odot}$ respectively.

For the longer period solution, the black hole case is even stronger, with 
$f(M)=6.60 \pm 
0.36 M_{\odot}$, $q=0.12 \pm 0.02$, $M_1 \sin^{3} i = 8.28 \pm 
0.54 M_{\odot}$, $M_1 \ge 8.95 \pm 0.58 M_{\odot}$ and $M_2 \ge 1.07 \pm 
0.19 M_{\odot}$. Both the long period and large mass function are 
reminiscent of the black hole transient V404 Cyg (Casares, Charles \& Naylor 
1992).

\section{The Nature of the Companion Star and distance to BW Cir}

In order to refine the spectral classification we have averaged our
individual spectra in the rest frame of the companion star, using our
first orbital solution, and compared it with our optimally broadened
spectral type templates in the range F5-K7III using a $\chi^2$
minimization routine (see Marsh, Robinson \& Wood 1996 for
details). The minimum $\chi^2$ is obtained for spectral types
G0III-G5III, which contribute 39-35 \% to the optical flux in our
spectral range (see table 1). Figure 3 presents the averaged spectrum
of BW Cir in the companion's rest frame, together with our G5III
template\footnote{HD 62351 is a close binary with a separation of 0.4", 
resolvable through speckle. It is classified in the V-band as a G5III or 
G5/6IV (Christy \& Walker 1969, Houk \& Smith-Moore 1988). We confirm this 
classification in the R-band based on the line ratio 
$H_{\alpha}$/$\lambda6498$, which is very sensitive to $T_{\rm eff}$ 
(see table 1).}, broadened by $V \sin i = 71$ km s$^{-1}$. 

\begin{figure}
\includegraphics[angle=-90, scale=0.33]{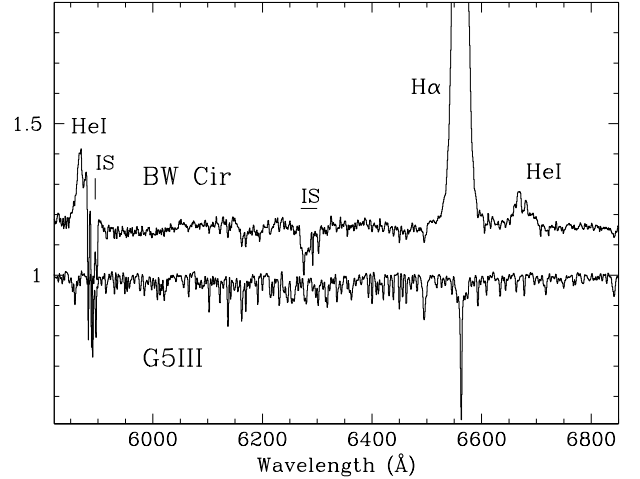}
\caption{Doppler corrected average of the 57 individual BW Cir spectra in the
rest frame of the companion star and the G5III template, broadened by 71 km
s$^{-1}$. Note the extreme weakness of the metallic features from the 
companion star, as compared to the template.  
\label{fig3}}
\end{figure}

A normal G0III-G5III typically has $M_2=2.1-2.4 M_{\odot}$ and $R
\simeq 5-8 R_{\odot}$ (Gray 1992). However, the Roche lobe
equivalent radius of a $2.1-2.4 M_{\odot}$ star in a 2.5d orbit
is $\simeq 4.6-4.8 R_{\odot}$ and hence the companion is probably a lower 
mass, but more evolved star. 
Since $T_{eff}$ is constrained by the spectral 
classification ($\simeq$ 5000 K), and the radius of the companion by the 
Roche lobe geometry\footnote{This results from combining Kepler's Third 
Law with Paczy\'nski's relation (Paczy\'nski 1971).} 

\begin{equation}
\left(\frac{R_{\rm L_2}}{\rm R_{\odot}}\right) 
\simeq 0.234 \left(\frac{P^{2/3}}{hrs}\right) 
\left(\frac{M_{\rm 2}^{1/3}}{M_{\odot}}\right),
\end{equation}

\noindent
we can apply the Stefan-Boltzmann relation to constrain the distance 
to BW Cir. Our orbital solutions yield $M_2 \ge 1.02$ M$_{\odot}$ and hence  
$R_2 \ge 3.6$ R$_{\odot}$. Therefore the Stefan-Boltzmann relation gives 
$M_{bol} \le 2.6$ which, combined with the bolometric correction and 
(V-R) colour for a G5III star (Gray 1992), gives $M_R \le 2.11$. The 
dereddened magnitude of the companion star can be estimated from the
observed quiescent magnitude (R=20.5), corrected for reddening ($E_{B-V}\sim 
1$; Kitamoto et al. 1990) and our determined 65 \% veiling. These yield 
$R \sim 19.3$. Therefore, the distance modulus relation provides 
a lower limit to the distance of 27 kpc which is substantially larger than 
the 10 kpc value proposed by Kitamoto et al. (1990).  Even so, we note that 
this makes $L_{\rm x} (1-10 keV) = 3.5 \times 10^{37} (d/10 kpc)^2 \ge 
2.6 \times 10^{38}$ erg s$^{-1}$ at the peak of the 1987 outburst, still 
confidently below the Eddington luminosity for a 10 M$_{\odot}$ black hole.

The distance has important implications for the observed systemic velocity. 
We have calculated the expected radial velocities of BW Cir, due to Galactic 
differential rotation, for a distance of 10 kpc and 27 kpc and these are -6  
and 104 km s$^{-1}$ respectively (Nakanishi \& Sofue 2003). The large distance 
scenario enables us to explain the observed $\gamma$-velocity without 
invoking any kick during the formation of the black hole. Although this 
scenario is the most attractive, we note the uncertainties involved in the 
IS extinction value and this result 
requires careful consideration. 
Our spectrum shows several IS absorption features which can be used to obtain 
an independent estimate of the reddening. The strongest ones are the NaI doublet 
for which we measure EW(D2)=1.37$\pm$ 0.04 \AA~ and EW(D1)=1.66$\pm$0.04 \AA. 
The empirical
calibration of Bardon et  al. (1990) yields $E_{B-V} = 0.25 \times EW = 0.78$ 
but this should be regarded as a lower limit because NaI(D2) is somewhat 
diluted by the broad HeI $\lambda$5875 emission from the accretion disc. 
We have also measured the EW of the diffuse IS band at $\lambda$6203 and find 
0.24 $\pm$0.02 \AA. Using the empirical calibration between EW and reddening 
from Herbig (1975) we get $E_{B-V} = 0.84 \pm 0.10$. These numbers are 
consistent with Kitamoto et al.'s reddening of $E_{B-V} \sim 1$, which was 
estimated by comparing the observed outburst colour of the optical counterpart 
with typical values of luminous LMXB given in van Paradijs (1983). This assumed 
that the X-ray irradiated disc dominated the optical emission, which is
confirmed by our results here and hence 
give further support to our estimated distance of $\ge$27 kpc. We note 
that this distance argues against BW Cir and Cen X-2 being the same 
object since it would make the 1966 outburst substantially super-Eddington 
(by more than an order of magnitude) for a 10-20 M$_{\odot}$ black hole. 
However, it does make BW Cir similar to the black hole 
LMXB GX339-4, where analysis of the NaI IS aborption features at high 
resolution has revealed complex velocity structures, consistent with a large 
distance of  $> 15$ kpc (Hynes et al. 2004). We also note that this implies  
$L_{\rm x} \sim 0.1~L_{\rm Edd}$ for the 1987 hard X-ray outburst which makes 
the most luminous hard state yet seen in a BH transient. 

\begin{figure}
\includegraphics[angle=-90, scale=0.33]{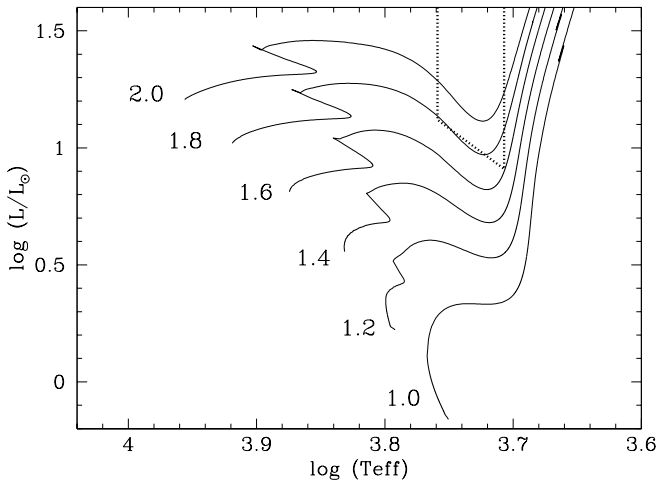}
\caption{H-R diagram with evolutionary tracks of solar metallicity stars 
of masses 1.0-2.0 M$_{\odot}$, from Cassini, Salaris \& Castelli (2004). 
Dotted lines mark the space region allowed by our spectral type 
classification ($T_{eff}=5091-5743$ K) and the Stefan-Boltzmann relation 
for the case of a Roche lobe filling star with $M_2 \ge 1.02$ M$_{\odot}$ 
and P=2.544 days. 
\label{fig4}}
\end{figure}

Additional constraints are provided by the position of the companion 
star in the H-R diagram, as compared with evolutionary sequences of single 
stars. Our spectral analysis indicates that the companion is well 
fitted by a G0-5III template and hence $T_{eff} \simeq 5091-5743$ K (Gray 1992). 
On the other hand, combining eq. 1 with the Stefan-Boltzmann relation gives 
$\log \left(\frac{L_2}{L_{\odot}}\right) \ge 4 \times \log T_{eff} - 13.92$,  
where we have used M$_2 \ge 1.02$ M$_{\odot}$. This defines a region 
in the H-R diagram (see Figure 4) to compare with evolutionary
tracks of 1-2 M$_{\odot}$ stars, computed using the code 
BASTI\footnote{Pietriferni et al. 2004, in preparation; 
http://www.te.astro.it/BASTI.} for the case of 
Population I stars of solar metallicity. 
The diagram shows that the companion star is located in the Hertzsprung gap or 
the lower part of the giant branch for $M_2 
\ge 1.6-1.8$ M$_{\odot}$. Since the luminosity of a Roche lobe filling star 
depends weakly on $M_2$, the companion cannot be much more massive than 
$\simeq 1.6 M_{\odot}$ (for instance $\log L_2=1.06~ L_{\odot}$ for $M_2=2.0 
M_{\odot}$). 
If it were a 1.6 M$_{\odot}$ star with $T_{eff}\sim 5000$ K, 
then $\log L_2=1.0~ L_{\odot}$ and the distance would be $\sim$ 33 kpc. 
However, we note the limitations of this crude estimate because the donor 
stars in X-ray binaries are not in thermal equilibrium and their H-R tracks 
may deviate significantly from single star evolution.

\acknowledgments

JC acknowledges support by the Spanish MCYT grant AYA2002-0036.
TS acknowledges support by the programme Ram\'{o}n y Cajal.
MOLLY and DOPPLER software developed by T.R. Marsh is gratefully 
acknowledge. We thank an anonymous referee for helpful comments to 
the manuscript. Based on data collected at the European Southern 
Observatory, Monte Paranal, Chile.

\begin{deluxetable}{lccccccccc}
\tabletypesize{\scriptsize}
\tablecaption{Radial Velocity solutions and Spectral Type classification 
\label{tbl-1}}
\tablewidth{0pt}
\tablehead{
\colhead{Template} & \colhead{Spect. Type} &
\colhead{$H_{\alpha}$/$\lambda6498$} &\colhead{$\gamma$}  &  \colhead{P} &  
\colhead{To} & \colhead{$K_2$}  &  \colhead{$V \sin i$} & \colhead{f} & 
\colhead{$\chi^{2}_{\nu}$}\\
& & & \colhead{(km s$^{-1}$)} & \colhead{(days)} & \colhead{(+2453140.0)} & 
\colhead{(km s$^{-1}$)} & \colhead{(km s$^{-1}$)} & & \\
}
\startdata

HD 182901 & F5III & 8.2(1) & 103(3) & 2.54447(12) & 0.987(9) & 276.4(3.4) & 
66(4) & 0.47(2) & 2.50  \\
HD 83273  & G0III & 6.0(1) & 102(3) & 2.54445(9) & 0.983(8) & 279.3(3.2) & 
69(4) & 0.39(1) &   2.47 \\
HD 62351  & G5III & 3.7(1) & 103(3) & 2.54448(9) & 0.985(8) & 279.2(2.9) & 
71(4) & 0.35(1) &   2.42\\
HR 6915  & G8III & 2.7(4) & 104(2) & 2.54451(8) & 0.990(7) & 278.0(2.7) & 
71(4) & 0.25(1) & 2.56\\
HR 6925  & K3III & 1.6(1) & 109(2) & 2.54451(7) & 1.000(6) & 279.2(2.0) & 
59(4) & 0.15(1) &   2.58\\
HD 184722 & K7III & 1.3(1) & 110(2) & 2.54448(7) & 0.990(6) & 283.7(2.1) & 
61(4) & 0.13(1) &   2.66\\
\enddata
\end{deluxetable}


\begin{thebibliography}{}

\bibitem[]{401}Al-Naimy H., 1978, Astroph. Space Sci., 53, 181
\bibitem[]{402}Barbon R., Benetti S., Cappellaro E., Rossino L., Turatto M., 
1990, A\&A, 237, 79	
\bibitem[]{404}Brocksopp C., Jonker P.G., Fender R.P., Groot P.J., van der Klis 
M., Tingay T., 2001, MNRAS, 323, 517
\bibitem[]{}Casares J., Charles P.A., Naylor T. 1992, Nature, 355, 614
\bibitem[]{406}Charles, P.A. \& Coe, M.J. 2004, Compact Stellar X-Ray Sources, 
eds. W.H.G. Lewin \& M. van der Klis, CUP (astro-ph/0308020)
\bibitem[]{408}Gray, D.F. 1992, The Observation and Analysis of Stellar
Photospheres, CUP 20
\bibitem[]{}Christy, J.W. \& Walker, R.L. Jr., 1969, PASP, 81, 643
\bibitem[]{411}Herbig, G.H. 1975, ApJ, 196, 129
\bibitem[]{}Houk, N. \& Smith-Moore, M. 1988, Michigan Spectral Survey, Ann Arbor, 
Dep. Astron., Univ. Michigan, 4
\bibitem[]{412}Hynes, R.I., Steeghs, D., Casares, J., Charles, P.A. \& 
O'Brien K. 2003, ApJ, 583, L95
\bibitem[]{414} --------. 2004, ApJ, 609, 317
\bibitem[]{416}Ilovaisky, S.A., Pedersen, H. \& van der Klis, M. 1987, IAU 
Circ. 4362 
\bibitem[]{420}Kitamoto, S., Tsunemi, H., Pedersen, H., Ilovaisky, S.A. \& van 
der Klis, M. 1990, ApJ, 361, 590
\bibitem[]{422}Makino, F. et al. 1987, IAU Circ. 4342 
\bibitem[]{}Marsh, T.R., Robinson, E.L. \& Wood, J.H. 1994, MNRAS, 266, 137
\bibitem[]{423}Martin, A. 1995, PhD Thesis, Oxford Univ.
\bibitem[]{424}McClintock, J.E. \& Remillard, R. A. 2004, Compact Stellar 
X-Ray Sources, eds. W.H.G. Lewin \& M. van der Klis, CUP (astro-ph/0306213)
\bibitem[]{426}Nakanishi, H. \& Sofue, Y. 2003, PASJ, 55, 191
\bibitem[]{429}Paczy\'nski, B. 1971, ARA\&A, 9, 183
\bibitem[]{}Revnivtsev, M.G., Borozdin, K., Priedhorsky, W.C. \& Vikhlinin, 
A. 2000, ApJ, 530, 955
\bibitem[]{}Rhoades, C.E. \& Ruffini, R. 1974, Phys. Rev. Lett., 32, 324
\bibitem[]{432}van Paradijs, J. 1983, Accretion-driven Stellar X-Ray 
Sources, eds. W.H.G. Lewin \& E.P.J. van den Heuvel, CUP, p.207
\bibitem[]{434}van Paradijs, J. 1996, ApJ, 464, L139  
\bibitem[]{438}Wade, R.A. \& Horne, K. 1988, \apj, 324, 411


\end{thebibliography}
\end{document}